\documentclass[aps,prr,letterpaper,10pt,twocolumn,showpacs,superscriptaddress]{revtex4-1}
\usepackage[utf8]{inputenc}

\usepackage{amsmath}
\usepackage{amsfonts}
\usepackage{amssymb}
\usepackage{graphicx}
\usepackage{xcolor,ulem}

\usepackage[colorlinks=true,bookmarks=false,citecolor=blue,urlcolor=blue]{hyperref}

\newcommand{\Xzero}{x}
\newcommand{\szero}{s}
\newcommand{\Xone}{X}
\newcommand{\sone}{S}

\newcommand{\bigO}[1]{{\mathcal{O}({#1})}}

\renewcommand{\emph}{\textit}

\begin{document}

\title{Fast and simple qubit-based synchronization for quantum key distribution}
\author{L.~Calderaro}
\author{A.~Stanco}
\author{C.~Agnesi}
\author{M.~Avesani}
\affiliation{Dipartimento di Ingegneria dell'Informazione, Universit\`a degli Studi di Padova, via Gradenigo 6B, 35131 Padova, Italy}
\author{D.~Dequal}
\affiliation{Matera Laser Ranging Observatory, Agenzia Spaziale Italiana, Matera, Italy}
\author{P.~Villoresi}
\email{paolo.villoresi@dei.unipd.it}
\affiliation{Dipartimento di Ingegneria dell'Informazione, Universit\`a degli Studi di Padova, via Gradenigo 6B, 35131 Padova, Italy}
\author{G.~Vallone}
\affiliation{Dipartimento di Ingegneria dell'Informazione, Universit\`a degli Studi di Padova, via Gradenigo 6B, 35131 Padova, Italy}
\affiliation{Dipartimento di Fisica e Astronomia, Universit\`a degli Studi di Padova, via Marzolo 8, 35131 Padova, Italy}

\date{\today}

\begin{abstract}
    We propose Qubit4Sync, a synchronization method for Quantum Key Distribution (QKD) setups, based on the same qubits exchanged during the protocol and without requiring additional hardware other than the one necessary to prepare and measure the quantum states. Our approach introduces a new cross-correlation algorithm achieving the lowest computational complexity, to our knowledge, for high channel losses. We tested the robustness of our scheme in a real QKD implementation.
\end{abstract}

\maketitle

\section{Introduction}
Quantum Key Distribution (QKD) constitutes a promising technology for the security of the future communication networks. Introduced in 1984~\cite{Bennett1984}, QKD is a communication protocol for the generation of a secret key shared only by two parties, which afterwards can be used to establish a secure communication. The selling point of QKD is that the security of the protocol is guaranteed as long as the laws of Quantum Mechanics are valid. This is a great leap forwards compared to similar classical protocols which are based on the limited computational power of the adversary. The practical implementation of the protocol has developed to the point that several experiments have been performed exploiting deployed telecom fibers~\cite{Yoshino2013}, daylight free-space channel in urban areas~\cite{Avesani2019,Gong2018,Liao2017a}, satellite-to-ground channel~\cite{Bedington2017,Liao2017}. Nonetheless, there still remains several challenges to be addressed as communication rate and range, and making QKD systems low cost, compact and robust~\cite{Diamanti2016}.

Clock synchronization is crucial for communication networks~\cite{Bellamy1995,Bregni1997,Narula2018}, QKD not being an exception. Indeed, it is fundamental in QKD protocols not only because it allows the two parties to correctly generate the secret key, but also to filter out the noise. The knowledge of the time in which the signal is expected to arrive at the receiver allows to discard the majority of the detection due to noise, increasing the signal-to-noise ratio (SNR). This is of crucial importance as the SNR is usually the limiting factor for the performances of QKD. The solutions which are usually adopted in current QKD implementations include either to send a decimated copy of the transmitter's clock through a separated single-mode fiber~\cite{Korzh2015} or even the same quantum channel~\cite{Liu2010}, or to lock the two clocks to an external time reference provided for instance by GNSS receivers~\cite{Avesani2019,Vallone2015,Jennewein15}. All these solutions imply the use of additional hardware and hence an increase in complexity and cost of the setup.

In this work, we propose Qubit4Sync,  a synchronization system that only uses the same qubits exchanged during the QKD protocol, without requiring additional hardware. Our approach is to exploit the information on the measurements that the receiver performs on the qubits. Hence, a pre-analysis is performed before the standard QKD post-processing, extracting the information on the time of arrival of the signal.

\section{Description of the Algorithm}
Alice transmits a qubit string (the raw key) encoded in the state of a train of attenuated optical pulses. The time between two consecutive qubits, $\tau^A$, 
is set by Alice's clock. On the other side, Bob receives some of the qubits (due to losses), analyzes their state and uses his clock to measure their time of arrival. We consider the case in which Alice and Bob's clocks may have a time bias as well as a relative drift in time of their frequencies.
This implies that Bob may measure a different time $\tau^B$ between subsequent qubits.

The goal of Bob is to determine the position of the detected qubits in Alice's raw key: this operation is needed to correctly generate the sifted key, perform the parameter estimation and the subsequent post-processing.
The above problem can be reformulated as follows: Bob needs to determine the expected time of arrival (measured by his clock) of the qubits sent by Alice, namely he needs to solve two tasks:
\begin{itemize}
    \item [i)] {\it Period recovery}: to recover the period $\tau^B$ from the obtained detections. 
\item [ii)] {\it Time-offset recovery}: to determine the time delay
between the measured and sent sequence.
\end{itemize}
Step i) is needed to correctly reconstruct the separations in the raw key between consecutive detections.
Step ii) is needed to
associate each detection to the corresponding bits in Alice's raw string.

This problem can be solved by synchronizing Alice and Bob's clocks and by knowing the time of flight of the qubits \cite{Narula2018}. However, Bob just needs to know at which time Alice's pulses will arrive, and not at which time she sent them. Therefore, their clocks may be synchronous up to a time offset.

We define $t_a^{\rm m}$ as the measured time of arrival (according to Bob's clock) with
$a\geq 1$ enumerating the obtained detections.
Since the time separation between the qubits is constant at Alice site, 
a model that reproduces the expected time of arrival $t_a^{\rm e}$ at Bob site can be expressed as
\begin{equation} \label{eq:time_expected}
    t_a^{\rm e} = t_0 + n_a\tau^B+ \epsilon_a\,,\quad
    n_a\in\mathbb N
\end{equation}
The index $n_a$ identifies the position of the sent qubit in the raw key of Alice, $t_0$ is the expected time of arrival of the first pulse sent by Alice, while $\epsilon_a$ is a normal random variable with zero mean and variance $\sigma^2$ (due to the measurement jitter).
If Alice and Bob's clock are perfectly synchronized, then $\tau^B=\tau^A$.
We note that we are neglecting in the model the noise. 

We define the \textit{Time Error} function  
$\mathrm{TE}_a$ between measured and expected time of arrival as
\begin{equation}
    \mathrm{TE}_a = t_a^{\rm m} -
    t_a^{\rm e}
\end{equation}
The time error variation between two different detection $a$ and $a+b > a$
is the so called \textit{Time Interval Error} $\mathrm{TIE}_{a}(b)$, defined as \cite{Bregni1997}
\begin{equation}
    \mathrm{TIE}_{a}(b) = \mathrm{TE}_{a+b} - \mathrm{TE}_{a}
\end{equation}

Below we will describe how the above two tasks (frequency and time-offset recovery) can be realized by using only the qubits exchanged during the QKD protocol, without requiring additional hardware.

\subsection{Period recovery}
We first describe how the period $\tau^B$ can be obtained by Bob. 
Let's suppose that Bob acquires data for a time $T_{\rm acq}$ and in this time the relative frequencies of Alice and Bob's clock are constant, namely the periods $\tau^A$ and $\tau^B$ are constant. Typical values of 
$T_{\rm acq}$ are of the order of 1 sec.
The above acquisition corresponds to $M$ pulses sent by Alice with $M=\lfloor T_{\rm acq}/\tau^B\rfloor $, of which $D$ are the one detected by Bob. We can label the detected pulses with the index $b = 1,\ldots, D$ such that $t^m_{a+b}-t^m_a<T_{\rm acq}$, being $t^m_a$ the last detection before the acquisition started.
We define the condition of successful period recover when the following condition is satisfied:
\begin{equation} \label{eq:TIE_bound}
    |\mathrm{TIE}_{a}(b)| < \frac{\tau^B}{2}\quad {\rm for}\,{\rm all}\quad b=1,\ldots,D
\end{equation}
The latter condition implies that the $M$ subsequent pulses sent by Alice during the acquisition correspond to exactly $M$ time-slots of length $\tau^B$ on Bob's clock. 
We note that eq.~\eqref{eq:TIE_bound} is a sufficient condition to correctly reconstruct the separations, $n_{a+b} - n_a$, in the raw key between consecutive detections, but it is not the optimal one as the signal may arrive at any time inside the time slot $\tau^B$. This would prevent to filter out the noise. The optimal value for $\tau^B$ is the one such that
\begin{equation} \label{eq:TIE_bound_tight}
    \frac1D\sum_{b=1}^{D}|\mathrm{TIE}_{a}(b)|^2 \simeq  \sigma^2
\end{equation}

To have a first guess $\tau_{0}^B$ about the value of $\tau^B$, Bob should performs a fourier analysis of the times of arrival signal~\cite{FFTW05}. 
The latter is a sequence of $N$ symbols (taking value 0 or 1), with the ones corresponding to the times of detections.
Assuming that Alice's clock frequency is less than twice the one of Bob, we may sample the time of arrival of the photons with $4/\tau^A$ sampling rate (since the sample is real-valued half of the spectral range of the DFT is meaningful). 
For the purpose of real-time analysis (i.e. to speed up computation), we perform the fast fourier transform (FFT) limiting the number of samples to $N=10^6$ \cite{FFTW05}, namely we limit the sampling time for the FFT to $T_{\rm samp}=N(\tau^A/4)$.

The above FFT will provide an estimate $\tau_{0}^B$ of $\tau^B$ with an error of $\sim 4 \tau^A/N$. We note that $\tau_{0}^B$ satisfies  eq.~\eqref{eq:TIE_bound} for the first $b=1,\ldots,D_0$ detections, such that $t^m_{a+b}-t^m_a<T_{\rm samp}$. However, if the acquisition time $T_{\rm acq}$ is larger than $T_{\rm samp}$ (i.e. $M>N/4$), the estimate $\tau_{0}^B$ may not be sufficiently accurate and the condition eq.~\eqref{eq:TIE_bound} could be not satisfied.

Instead of performing a fourier transform of $4M$ samples (that could increase computational complexity), we perform a linear regression of $\mod_{\tau^B_0}(t^m_{a+b})$ as a function of the measured time $t^m_{a+b}$, for $b=1,\ldots,D_0$. We use a least trimmed squares algorithm as a robust statistical method against background~\cite{Li2005}. While the intercept does not provide any useful information, it is easy to prove that the slope of the linear model is equal to $(\tau^B - \tau_{0}^B)/\tau^B$, with which we have an estimate of $\tau^B$ such that eq.~\eqref{eq:TIE_bound_tight} is satisfied. 

Once  $\tau^B$ has been identified, Bob can associate each detection to a different slot of length $\tau^B$, indicated by the indices $n_{a+b}$, up to a constant (depending on $t^e_0$). Indeed, Bob can calculate all the index differences by using the relation $n_{a+b}-n_{a}=[\frac{t^{\rm m}_{a+b}-t^{\rm m}_{a}}{\tau^B}]$. 

We note that the variation of $\tau^B$ during a given  acquisition time $T_{\rm acq}$ should be small in order to guarantee that eq. \eqref{eq:TIE_bound_tight} could be satisfied, namely $M|\delta\tau^B|=|\frac{\partial \tau^B}{\partial t}|\frac{T^2_{\rm acq}}{\tau}\lesssim10\sigma$. If the latter condition is not satisfied, the period recovery should be performed by reducing the acquisition time $T_{\rm acq}$.

In the next acquisition of $D'$ pulses, the value of $\tau^B$ may change due to a relative frequency shift of the clocks. The condition becomes $\sum_{b=1}^{D'}|\mathrm{TIE}_{a+D}(b)|^2 \simeq D' \sigma^2$, that will be satisfied applying again the above analysis.

\subsection{Time-offset recovery}
We now describe how the initial delay $t_0$ can be estimated, allowing to determine the index $n_a$'s. We restate that once the period recovery has been performed, Bob has correctly estimated $\tau^B$ and the index differences $n_{a+1}-n_{a}$ for $a\geq 1$. Then, only the first index $n_1$ is needed to calculate the set of indices $\{n_{a}\}$. We note that $n_1$ is related with $t_0$ by $t_0=t^{\rm e}_1-n_1\tau^B$.

Due to losses in the channel, with high probability the first pulse will be not detected by Bob. Moreover, the presence of background makes it not straightforward to distinguish the detection from Alice's qubit.

As a first guess, we may identify the first Bob detection as the first pulse sent by Alice (i.e. $n_1=0$). The first detection can be identified by looking for a rising edge of the detection frequency. If the overall transmittance of the system is $\eta$ and the mean number of photons per pulse sent by Alice is $\mu \sim 1$, Bob expects to have a detection each $1/\eta$ pulses. Therefore, the uncertainty on $n_1$ will be of the order of $1/\eta$.

To precisely determine $t_0^A$, our approach is to to calculate the correlation between the signal received by Bob with a synchronization string $s^A$ that has length $L\gg 1/\eta$. The string $s^A$, which is also known to Bob, is placed at the beginning of Alice's raw string. We encode $s^A$ in the base which is more frequently measured by Bob (say the $Z$ basis) and we assign the values $+1$ or $-1$ to the two orthogonal states of such basis. In case of no detection, we assign the value $0$. Then, once Bob has determined the period $\tau^B$ and has a first guess about $t_0$ (hence $n_1$), he can produce a string $s^B$ with values $0$, $-1$ or $+1$. In order to precisely determine $t_0$ we may exploit the cross-correlation between the signal received by Bob with $s^A$: indeed, the value that maximize the cross-correlation corresponds to the needed offset. 

We here recall that the cross-correlation function between $s^A$ and $s^B$
is defined as ($m=0,\cdots,L-1$):
\begin{equation} 
\label{eq:Xzero_auto}
    \Xzero_m^{AB} = \frac1L\sum_{n=0}^{L-1} s_{n+m}^{*A} s_{n}^{B}
    \end{equation}
with the convention that $s_{n'}^A=s_{n'-L}^A$ for $n'\geq L$.
The offset between Alice and Bob strings is:
\begin{equation} \label{eq:offset}
    \mathrm{TE}_0 / \tau^B =n_1= m_{\rm opt}
\end{equation}
where $m_{\rm opt}$ is the value of $m$ that maximize the
cross-correlation $\Xzero_m^{AB}$.
By exploiting the convolution theorem, the maximum of the cross-correlation $\Xzero_m^{AB}$ can be evaluated by Bob with 
$\bigO{L\log_2L}$ operations (we are assuming that the Fourier transform of $s^A$ is already known by Bob before the measurements). Below
we propose a new algorithm that reduces computational complexity.

Our method is based on the properties of particular synchronization strings, that allows to calculate the cross-correlation more efficiently. More precisely, 
we need to use a synchronization string such that its auto-correlation function $\Xzero_m^{AA}$ has $N_1$ periodic peaks, namely it
satisfies
\begin{equation}
    \begin{aligned}
    &\Xzero_0^{AA}=1
    \\
    &\Xzero_{jL_1}^{AA}\simeq c_0 && \text{for }j>0
    \\
    &\Xzero_{u+jL_1}^{AA}\simeq 0 &\quad& \text{for }u>0
    \end{aligned}
\end{equation}
where $0<c_0<1$,
$L = N_1 L_1$, $N_1$ and $L_1$ being integer numbers, $u=0,\cdots,L_1-1$ and $j=0,\cdots,N_1-1$.
The method to generate a string $\szero^A$ that satisfies eq.~\eqref{eq:Xzero_auto} 
is described in appendix. Fig.~\ref{fig:exampleX} shows the auto-correlation $\Xzero$ of such a string with $L=10^6$ and $N_1=10$.
We leave for future investigation,
the study of the optimal $c_0$ in function of losses and errors.
\begin{figure}[t]
    \centering
    \includegraphics[width=\columnwidth]{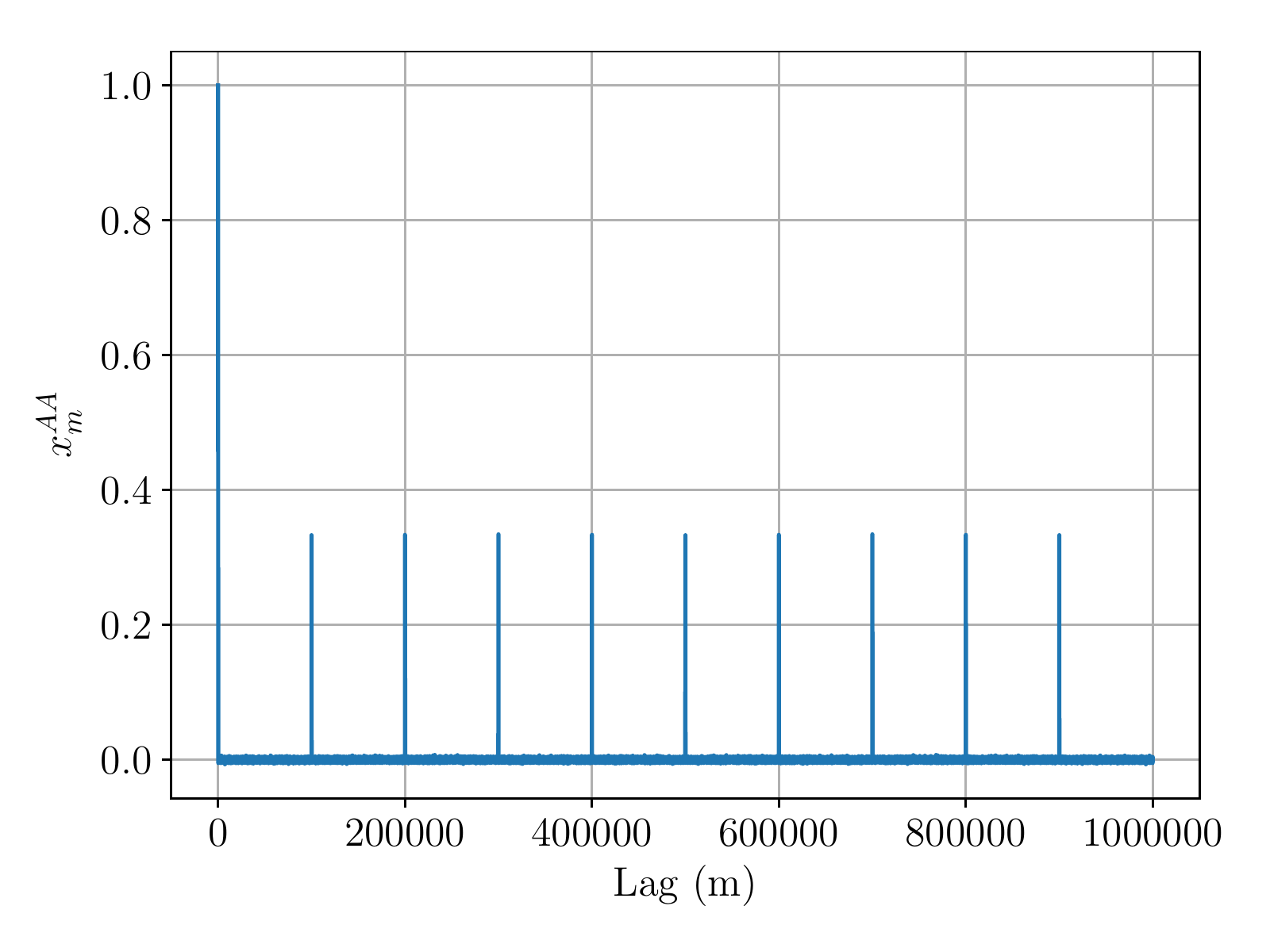}
    \caption{Example of auto-correlation $\Xzero^{AA}_m$ for a synchronization string with $L=10^6$, $N_1=10$ and $c_0=\frac13$.}
    \label{fig:exampleX}
\end{figure}

\begin{figure*}[t]
    \centering
    \includegraphics[width=\textwidth]{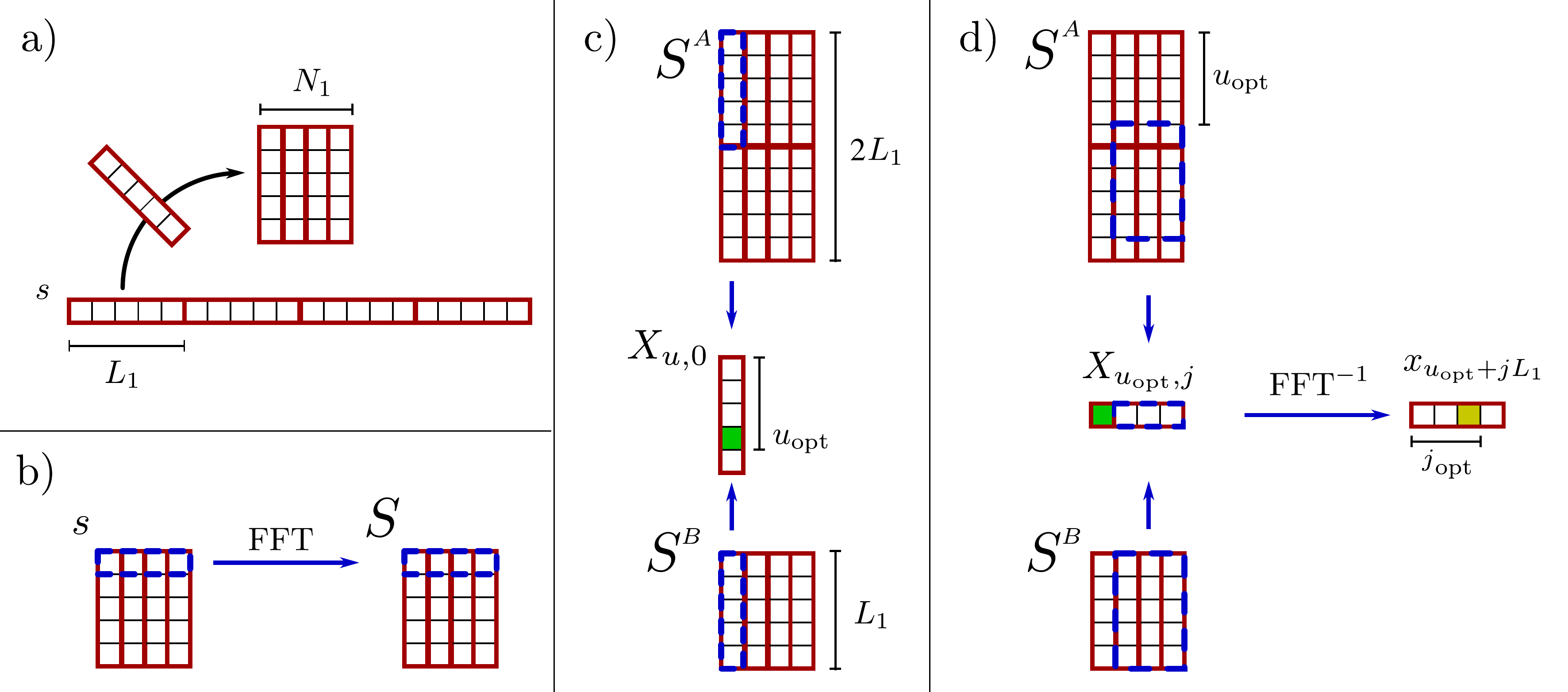}
    \caption{\textbf{a)} The string of Alice, $s^A$, and Bob, $s^B$, are divided in $N_1$ blocks of length $L_1$ and reshaped into a $L_1 \times N_1$ matrix. \textbf{b)} For each row of the matrix the FFT is calculated obtaining the matrices $\sone^A$ and $\sone^B$. Note that $\sone^A$ can be calculated in advance. \textbf{c)} The cross-correlation ${X}_{u,0}$ of the first columns of $\sone^A$ and $\sone^B$ is calculated. The position $u_\mathrm{opt}$ that maximizes $\hat{X}_{u,0}$ corresponds to the position of the first peak of the cross-correlation $X$. \textbf{d)} Consider the block of $S^A$ shifted by $u_{\rm opt}$ rows and calculate the cross-correlation between the remaining columns of $\sone^A$ and $\sone^B$. The resulting vector $\Xone^{AB}_{u_{\rm opt},j}$ is anti-transformed so to obtain $x^{AB}_{u_{\rm opt}+jL_1}$. The $j_\mathrm{opt}$ that maximize $x^{AB}_{u_{\rm opt}+jL_1}$ provides the position of the major peak among the smaller peaks.}
    \label{fig:alg_offset}
\end{figure*}

To simplify computational complexity we may exploit the periodicity of the auto-correlation. We need to first  calculate the interleaved sum of $x^{AB}_m$ defined as
$\frac1{N_1}
\sum_{j=0}^{N_1-1}\Xzero^{AB}_{u+jL_1}$.
To do so, we need to evaluate $\sone^A$ ($\sone^B$), the interleaved DFT (discrete fourier transform) of $\szero^A$ ($\szero^B$):
\begin{equation} \label{eq:sone}
\sone^A_{r,j} = 
\sum_{k=0}^{N_1-1} \szero^A_{r+kL_1} e^{- \frac{2\pi i}{N_1}jk },
\end{equation}
where $r = 0,1,\ldots,L_1-1$ and $j = 0,1,\ldots,N_1-1$. The index $j$ span through the frequency domain, but the time domain is still present due to the index $r$.
We note that the above operation corresponds to
reshaping the sequence $s$ into a $L_1\times N_1$ matrix and calculating the FFT for each row (see Fig. \ref{fig:alg_offset}).
Therefore, we can define a cross-correlation in the time domain of $\sone^A$ and $\sone^B$
for $u = 0,1,\ldots,L_1-1$:
\begin{equation} \label{eq:Xone}
    \Xone^{AB}_{u,j} = \frac1{L_1}\sum_{r=0}^{L_1-1} (\sone^A_{r+u,j})^* \sone^B_{r,j}
\end{equation}
We note that the Fourier coefficient $\sone_{r,j}$ are defined for $r=0,\cdots,L_1-1$.
However, by extending the original definition \eqref{eq:sone} it is possible to define them for larger values of $r$, by the recursive relation
$\sone_{r+L_1,_j}=\sone_{r,j} e^{\frac{2\pi i}{N_1}j}$.

In appendix, we prove the following

\textit{Lemma 1:
the cross-correlation $\Xone^{AB}_{u,j}$ is
related to the cross-correlation
$\Xzero^{AB}_{u+jL_1}$ by a DFT:}
\begin{equation}
\label{DFT_X}
\Xzero^{AB}_{u+jL_1}
=\sum_{k=0}^{N_1-1}
e^{-\frac{2\pi i}{N_1}jk}
\Xone_{u,k}
\end{equation}
The interleaved sum of $x^{AB}_m$ can be easily evaluated by using eq.~\eqref{DFT_X}:
\begin{equation}
\label{interleaved_x}
\frac1{N_1}
\sum_{j=0}^{N_1-1}\Xzero^{AB}_{u+jL_1}=\Xone_{u,0}
    =\frac1{L_1}\sum_{r=0}^{L_1-1} (\sone^A_{r+u,0})^* \sone^B_{r,0}
\end{equation}

By defining 
$m_{\rm opt}=u_\mathrm{opt} + j_\mathrm{opt} L_1$, we may first determine $u_\mathrm{opt}$ by maximizing $\Xone_{u,0}$.
Indeed, due to the periodicity of the autocorrelation, the correlation $\Xone_{u,0}$ presents a single peak for $u=u_\mathrm{opt}$, namely we will have $X_{u_{\rm opt},0}\simeq c_0+\frac{1-c_0}{N_1}$ while $X_{u\neq u_{\rm opt},0}\simeq 0$. Thus, $u_{\rm opt}$ is the index that maximizes the averaged cross-correlation $X_{u,0}$.

The above relation provides a method to find the position of the $N_1$ peaks of $\Xzero^{AB}_m$, that are located at positions $m=u_{\rm opt}+jL_1$. 
To find $j_{\rm opt}$, we can now can use equation 
\eqref{DFT_X} to calculate (and maximize)
the cross-correlation only in such $N_1$ equally separated points,
$x_{u_{\rm opt}+jL_1}^{AB}$ for
$j=0,\cdots, N_1-1$.

\textit{Computational complexity}. The algorithm to calculate the offset can be visualized in Fig.~\ref{fig:alg_offset}. Alice and Bob strings, $s^A$ and $s^B$, are rearranged into two matrices with $L_1$ rows and $N_1$ column. For each row the FFT is calculated to find the matrices $\sone^A$ and $\sone^B$. $\sone^A$ can be calculated in advance, hence we consider just the computational cost for Bob's string which amount to $\bigO{L_1 N_1 \log N_1}$. At this point, we apply eq.~\eqref{interleaved_x} and calculate the cross-correlation $\Xone_{u,0}$ between the first columns of $\sone^A$ and $\sone^B$. This operation can be carried out with the FFT, for a computational cost of $\bigO{L_1\log_2L_1}$. We then find the position $u_\mathrm{opt}$ that maximizes $\Xone_{u,0}$.

Then we evaluate $\Xone_{u_{\rm opt},j}$ by eq.~\eqref{eq:Xone} for $j = 1,\ldots,N_1-1$ ($\Xone_{u_{\rm opt},0}$ have been already calculated) with a computational cost of 
$\bigO{L}$. Finally, by using {\it Lemma 1}, a FFT calculate 
$\Xzero^{AB}_{u_{\rm opt}+jL_1}$ and its maximum with $\bigO{N_1 \log N_1}$ operations. To summarize, the overall computational cost is $\bigO{(L+N_1)\log N_1 + \frac{L}{N_1} \log\frac{L}{N_1}}$
that can be optimize by choosing $N_1=\log(L)$, resulting in
\begin{equation}
    \bigO{L\log(\log L)}.
\end{equation}

To our knowledge, this is the most efficient algorithm for finding the maximum cross-correlation between two strings. Compared to other algorithms \cite{Hassanieh2012, Soliman2017, ZHAO2016}, the better efficiency comes with the disadvantage of a synchronization string satisfying eq.~\eqref{eq:Xzero_auto}. Therefore, this approach cannot be applied to pseudo-random strings. 

We note that our protocol shares some steps of the QuickSynch algorithm proposed in~\cite{Hassanieh2012}.
In particular, a similar method to obtain $u_{\rm opt}$ is used in~\cite{Hassanieh2012}. However, since in~\cite{Hassanieh2012} a pseudo-random string $s^A$ is used, the autocorrelation has a single peak and $X_{u_{\rm opt}}\simeq 1/N_1$: this is why in~\cite{Hassanieh2012} is suggested that $N_1$ repetitions of the $s^B$ string should be collected
to be able to determine the peak of $X_{u,0}$. 
Moreover, in~\cite{Hassanieh2012}, to estimate $j_{\rm opt}$, the correlation $\Xzero^{AB}_{u_{\rm opt}+jL_1}$ is estimated  by summing only $L_1$ points (namely they calculate, for all $j$'s, the quantity $\widetilde\Xzero^{AB}_{u_{\rm opt}+jL_1}=\frac{1}{L_1}\sum_{r=0}^{L_1-1} s^A_{r+u_{\rm opt}+jL_1}s^B_r$), while our method exploit relation \eqref{DFT_X} to calculate it efficiently and exactly.

\section{Experiment and results}

\begin{figure}[t]
    \centering
    \includegraphics[width=\columnwidth]{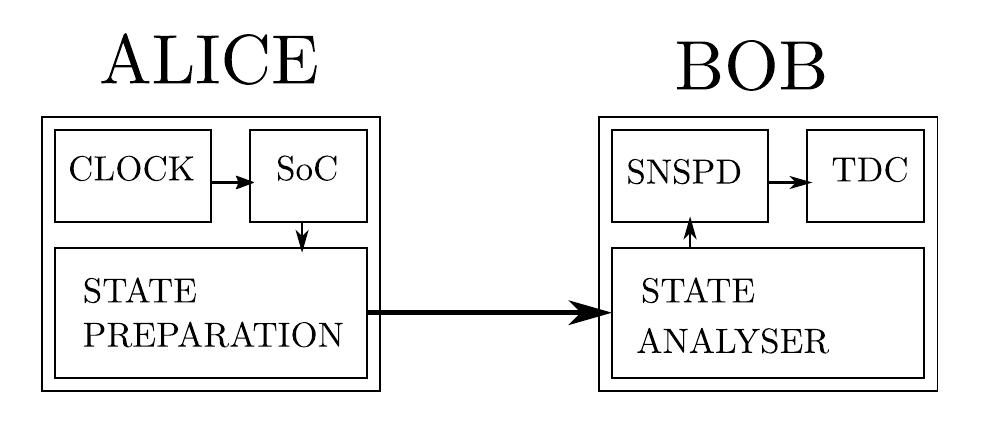}
    \caption{Setup}
    \label{fig:setup}
\end{figure}
We tested the Qubit4Sync algorithm in a QKD setup, illustrated in Fig.~\ref{fig:setup}. The quantum states are encoded in the polarization of attenuated laser pulses. Their polarization is modulated by a POGNAC source~\cite{Agnesi2019}, controlled by a Zynq-7000 ARM/FPGA System on a Chip (SoC, manufactactured by Xilinx). The time reference of Alice is given by a $10$~MHz reference signal to which the FPGA is locked. The repetition rate of the train of pulses is $50$~MHz, with a period of $\tau^A=20$~ns in  Alice's time. At the receiver side, a passive state analyzer performs the measurement on the polarization and four SNSPD detectors generate an electrical signal by the arrival of the optical pulse. A time-to-digital converter (TDC) measures the time of arrival, with $81$~ps time resolution. We do not provide any external time reference to the TDC, but its own internal clock. Then, the software processes the times of arrival every $T_{acq}=1$~s of acquisition time, analysing the frequency of the qubits. The offset analysis is performed just once with the data of the first second of acquisition.

\begin{figure}[t]
    \centering
    \includegraphics[width=\columnwidth]{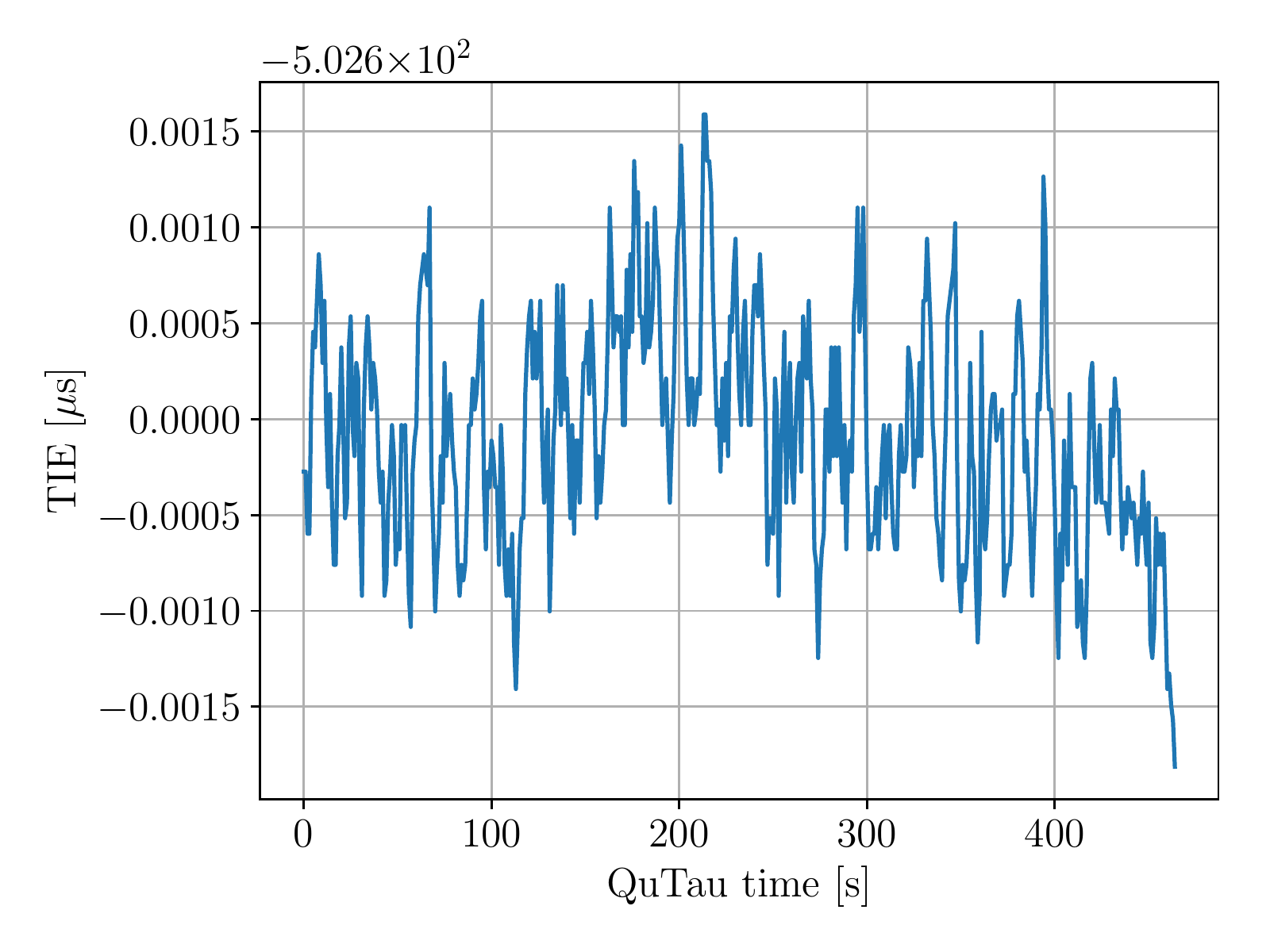}
    \caption{TIE between Alice and Bob's clock, after an interval of $1$~s, without Bob changing its clock pace (i.e. using $\tau^B=20$~ns in eq.~\eqref{eq:time_expected}).}
    \label{fig:TIE}
\end{figure}

Fig.~\ref{fig:TIE} shows the TIE, the time error between Alice and Bob's clocks after an interval of $T_{acq}$, in the case in which Bob is not correcting its clock (i.e. using $\tau^B=20$~ns in eq.~\eqref{eq:time_expected}). The graph shows that Alice and Bob's clocks accumulate a mean time error of about $0.5$~ms every interval of one second. This violates eq.~\eqref{eq:TIE_bound} as we have $\mathrm{TIE}_a \gg \tau^B$ and hence, if a period analysis is not performed, a correct separation in the raw key between consecutive detection cannot be achieved. We note that, with such TIE, a correct separation could be achieved only if Alice would be sending the pulses with $\tau^A > 0.5$~ms. Despite the large mean time interval error, the fluctuation around the mean is limited by $2$~ns over $400$~s.

We implemented the three polarization states version of the efficient BB84 \cite{Grunenfelder2018}, in which the receiver measures the polarization on the $Z$ and $X$ basis with $90\%$ and $10\%$ probability, respectively. The synchronization string, $\szero^A$, sent by Alice is entirely encoded in the $Z$ basis, so the $90\%$ of it will be decoded in the right basis (sifted).
For the purpose of the synchronization algorithm, just the number of sifted bits at Bob side matters. Hence, we will talk about overall transmittance $\eta$ as the ratio between the number of sifted bits at Bob side and the number of pulses sent by Alice.
The string sent by Alice is composed by a synchronization string, followed by random bits obtained from the quantum random number generator described in~\cite{Avesani2018}. We choose a number of states in the synchronization string $\szero^A$ of $L=10^6$, divided in $N_1=10$ blocks. If $\eta$ is the overall transmittance, the number of synchronization states received by Bob is $L\eta$. Therefore, assuming zero QBER and background noise, the maximum correlation value will be $\simeq\eta$, while the standard deviation of the correlation for other lags will be $\simeq\sqrt{\eta/L}$. The distinguishability, $\Delta$, of the maximum correlation peak among the others is given by the ratio of the former and the latter $\Delta \simeq \sqrt{L\eta}$. We set a threshold on the distinguishability of $\Delta \geq 10$, as successful detection of the maximum correlation. Hence, for our choice of $L$, the algorithm can cope with overall losses up to $40$~dB. In practice, the presence of background and misalignment between the transmitter and the receiver lowers the maximum losses that the algorithm can handle. 

\begin{figure}[t]
    \centering
    \includegraphics[width=\columnwidth]{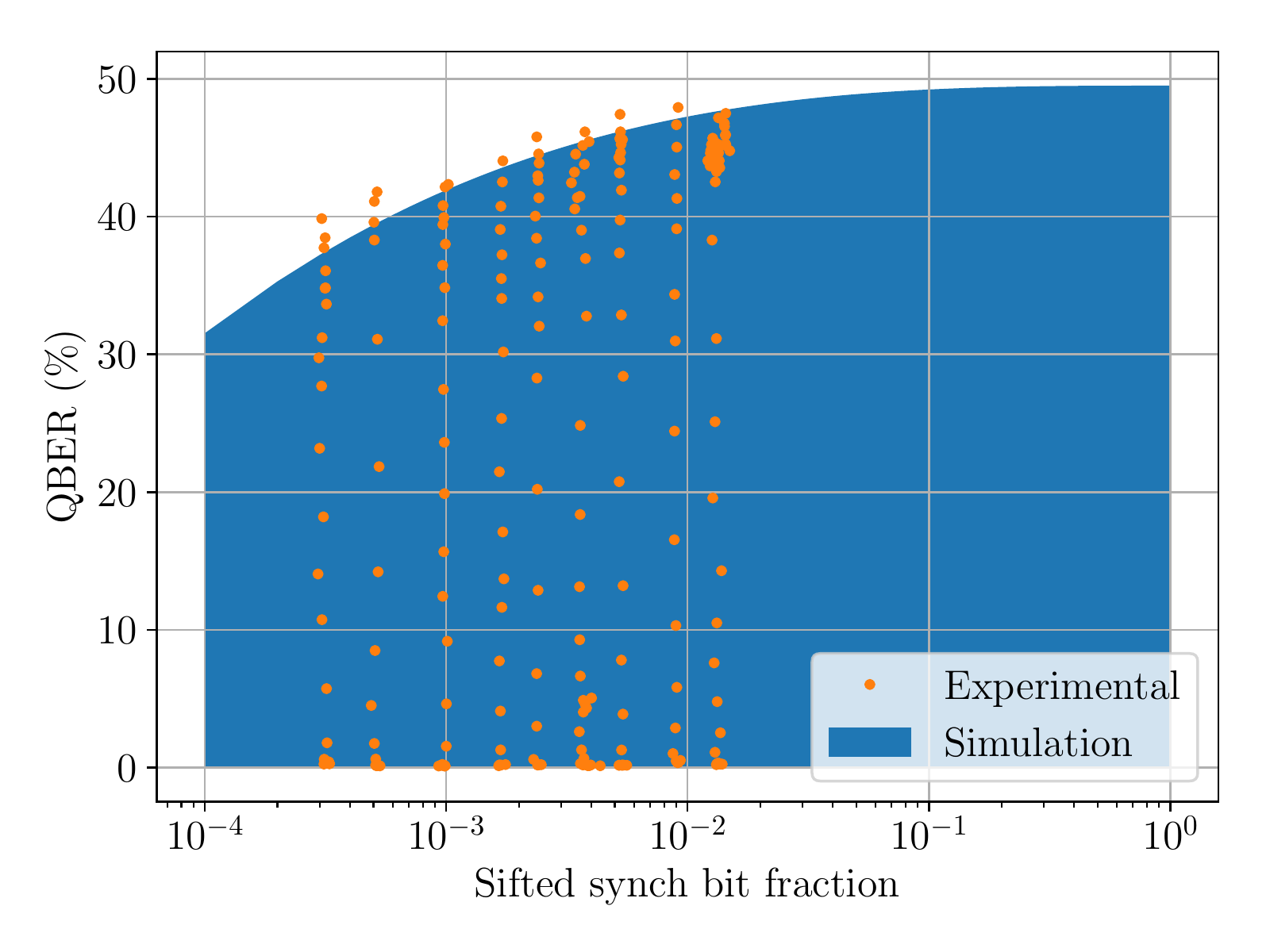}
    \caption{Successful synchronization for different values of QBER and detected bits. The blue region shows where the synchronization have been established using simulated data. Orange dots correspond to successful synchronization with data generated by our setup.}
    \label{fig:succ_synch}
\end{figure}

We tested the robustness of the offset analysis by tuning the QBER and the number of bits of $\szero^B$. We used strings generated from several QKD runs as well as simulations of the experiment. In particular, the simulation takes into account the losses and misalignment of the setup but not the presence of the background and dark counts. In Fig.~\ref{fig:succ_synch}, the result of the simulation is highlighted by the blue region, corresponding to the values of QBER and bits in $\szero^B$ in which the algorithm is expected to work. As regards the strings generated by the QKD setup, the orange dots show when the analysis was successful.

As expected, the simulation shows a good outcome of the analysis up to $10^{-4}$ sifted synchronization bit fraction. This is no longer true for high value of the QBER. Over $30\%$ of QBER the algorithm needs more bits in $\szero^B$ to contrast the reduction of the maximum correlation due to the bits flip. The background detection comes into play in the experimental runs, reducing the amount of losses the algorithm can tolerate. In our case, the analysis fails below a sifted synchronization bits ratio of $3\cdot 10^{-4}$, with 200~Hz of free-running background detection rate. The robustness to the QBER is comparable to what obtained with the simulated strings. The comparison is limited to a ratio of about $3\cdot 10^{-2}$ due to the maximum event rate our TDC can process.
It is interesting to note the very high robustness to the QBER, well above the threshold to establish a secure channel. In fact, a very rough alignment between transmitter and receiver is sufficient for the synchronization to take place. This implies that the precise alignment of the receiver and transmitter may be realized after the synchronization phase, maybe using the same states sent by Alice and without the use of external references that require additional lasers and detectors.

\section{Conclusions}
We have introduced Qubit4Sync, a new synchronization procedure only requiring the same photons encoding the quantum states that are exchanged in a QKD protocol. Moreover, we developed the fastest cross-correlation algorithm up to our knowledge. The common solution to synchronize two terminals in a QKD setup includes either an additional pulsed laser or a two GPS receivers. This work simplifies the practical implementation of a QKD setup because it avoids the use of additional hardware required for a synchronization sub-system, meaning cheaper apparatus and less failure probability due to hardware. 

Even though our procedure uses the qubits exchanged in the QKD protocol, the security is not undermined or weakened. The shared synchronization string is not used as part of the secure key, whereas the frequency analysis just uses the information on the time of arrival and not the one on the qubit state. The synchronization algorithm is also robust against eavesdropper's denial-of-service attack, since the QKD fails before the synchronization. Indeed, if an adversary tries to intercept the qubits the QKD protocol will stop when the QBER is above 11$\%$ \cite{Bennett1984, Grunenfelder2018}.

Our cross-correlation algorithm may be applied to GPS receivers, whose task is to correlate the signal sent by the satellite so to lock to its clock.

\appendix
\section{Method for the generation of the synchronization string}
We use the following method to generate a string $\szero$ that satisfies eq.~\eqref{eq:Xzero_auto}. From a uniform distribution in the $[-1,1)$ interval, extract $L_1$ real numbers $x_u$, with $u=0,\ldots, L_1-1$, and $L$ real numbers $y_{u,j}$, with $j=0,\ldots, N_1-1$. The synchronization string will take values as follows
\begin{equation}
    \szero_{u+jL_1} = 2\Theta(y_{u,j} - \lambda x_u)-1,
\end{equation}
where $\Theta$ is the Heaviside function and $\lambda$ a positive real value. The parameter $\lambda$ can be used to tune the value of $c_0$. Indeed, if $\lambda\leq 1$ we have $c_0=\frac{\lambda^2}{3}$, while if $\lambda>1$ then $c_0=1-\frac{2}{3\lambda}$. Fig.~\ref{fig:exampleX} shows the cross-correlation $\Xzero$ of such a string with $\lambda=1$, $L=10^6$ and $N_1=10$.

\section{Proof of Lemma 1}
The Fourier coefficient $\sone_{r,j}$ are defined
for $r=0,\cdots,L_1-1$.
However, from the original definition it is possible to extend their evaluation for larger values of $r$. Indeed, we may define
\begin{equation}
\label{extension}
\sone_{r+L_1,j}=\sone_{r,j} e^{\frac{2\pi i}{N_1}j}
\end{equation}
The above definition follows directly from eq.~\eqref{eq:sone}. The correlation can now be written as
\begin{align}\notag
&\Xzero^{AB}_{u+jL_1}
% =\frac1{L}\sum_{n=0}^{L-1}
% s^{A*}_{n+u+jL_1}
% s^{B}_{n}
=\frac1{L}
\sum_{k=0}^{N_1-1}
\sum_{r=0}^{L_1-1}
s^{A*}_{r+u+(k+j)L_1}
s^{B}_{r+kL_1}
\\\notag
&=\frac1{L}
\sum_{k=0}^{N_1-1}
\left[\sum_{r=0}^{L_1-u-1}
s^{A*}_{r+u+(k+j)L_1}
s^{B}_{r+kL_1}
+\right.
\\\notag
&\quad\left.\sum_{r=L_1-u}^{L_1-1}
s^{A*}_{r+u-L_1+(k+j+1)L_1}
s^{B}_{r+kL_1}
\right]
\end{align}
By using the definition of $S$ we obtain
\begin{align}\notag
\Xzero^{AB}_{u+jL_1}&
=\frac1{L}
\sum_{k,\ell_1,\ell_2=0}^{N_1-1}
\left[\sum_{r=0}^{L_1-u-1}
\sone^{A*}_{r+u,\ell_1}
\sone^{B}_{r,\ell_2}
e^{\frac{-2\pi i}{N_1}[(k+j)\ell_1-k\ell_2]}
\right.
\\\notag
&\quad\left.
+\sum_{r=L_1-u}^{L_1-1}
\sone^{A*}_{r+u-L_1,\ell_1}
\sone^{B}_{r,\ell_2}
e^{\frac{-2\pi i}{N_1}[(k+j+1)\ell_1-k\ell_2]}
\right]
\end{align}
By using the definition \eqref{extension},
for which we have
$\sone^{A}_{r+u,\ell_1}=\sone^{A}_{r+u-L_1,\ell_1}  e^{\frac{2\pi i}{N_1}\ell_1}$, if $r+u\geq L_1$ we obtain
\begin{align}
\notag
\Xzero^{AB}_{u+jL_1}
&=\frac1{L}
\sum_{k,\ell_1,\ell_2=0}^{N_1-1}
\sum_{r=0}^{L_1-1}
\sone^{A*}_{r+u,\ell_1}
\sone^{B}_{r,\ell_2}
e^{\frac{-2\pi i}{N_1}[(k+j)\ell_1-k\ell_2]}
\\\notag
&=
\sum_{\ell_1,\ell_2=0}^{N_1-1}
\frac1{L_1}
\sum_{r=0}^{L_1-1}
(\sone^{A}_{r+u,\ell_1})^*
\sone^{B}_{r,\ell_2}
e^{-\frac{2\pi i}{N_1}j\ell_1}
\delta_{\ell_1,\ell_2}
\\\notag
&=
\sum_{k=0}^{N_1-1}
e^{-\frac{2\pi i}{N_1}jk}
\left[\frac1{L_1}
\sum_{r=0}^{L_1-1}
(\sone^{A}_{r+u,k})^*
\sone^{B}_{r,k}
\right]
\end{align}
Finally, from the definition \eqref{eq:Xone}, we have the lemma:
\begin{equation}
    \Xzero^{AB}_{u+jL_1} = \sum_{k=0}^{N_1-1} e^{-\frac{2\pi i}{N_1}jk} \Xone_{u,k}^{AB}.
\end{equation}
The inverse relation is:
\begin{equation}
\Xone_{u,k}^{AB}
=\frac1{N_1}\sum_{j=0}^{N_1-1}e^{\frac{2\pi i}{N_1}jk}
\Xzero^{AB}_{u+jL_1}
\end{equation}
from which
\eqref{interleaved_x} can be
directly derived.

\bibliography{library}

\end{document}